\documentstyle[twocolumn,aps,prl]{revtex}

\begin{document}



{\bf Comment on ``Continuous quantum measurement: inelastic
tunneling and lack of current oscillations''}

\vspace{1ex}

Recent Letter \cite{b1} revisits the problem of continuous weak
measurement of quantum coherent oscillations of a qubit with a
linear detector, the role of which is played by the quantum point
contact (QPC) in the low-transparency regime. Previous studies of
this problem \cite{b2,b3,b4,b5} demonstrated that the qubit
oscillations with the Rabi frequency $\Omega$ should be reflected
in the spectral density $S(\omega)$ of the detector output
(electron current in the case of QPC detector) as a peak at
frequency $\omega =\Omega$. An interesting feature of such a 
peak \cite{b2} is that its height provides the measure of detector
``ideality'', i.e.\ shows how close the detector is to being
quantum-limited. For the quantum-limited detector the
maximum peak can reach four times the noise pedestal: $S(\Omega
)-S_0\leq 4S_0$, where $S_0$ is the output noise of the detector
(current shot noise for the QPC detector). This conclusion of
Ref.\ \cite{b2} has been confirmed in \cite{b3,b4,b5}. Similar
results have been also obtained for continuous measurement of a
single spin precession \cite{b7,b8}. Analogous spectral peak
associated with the Rabi oscillations of the
continuously-monitored qubit has been observed experimentally in
\cite{b9}.

In spite of all these developments, Letter \cite{b1} claims 
to show that there is no peak in the spectral density of
current in continuous qubit measurement by the QPC detector. 
The purpose of this Comment is to emphasize that this claim is
incorrect, and to point out the error in the arguments of 
\cite{b1}. The error is rather simple and stems from incorrect 
use of the ``conditional'' description of the measurement 
process. Conditional approach allows one to simulate individual 
random outcomes of a quantum measurement and has several 
important advantages, providing, e.g., a simple description of 
the feed-back control of quantum systems. However, it uses an 
{\em assumption} that quantum interference between different 
possible outcomes of measurement is suppressed, and may lead 
to incorrect conclusions if the basis for which the interference 
can be neglected is chosen incorrectly. Mistake of this type was 
made in Letter \cite{b1} which {\em assumes} that the qubit 
interaction with the QPC detector suppresses quantum interference 
between qubit energy eigenstates. While these states indeed 
decohere on the long time scales of order $\Gamma^{-1}$,
where $\Gamma$ is the qubit decoherence rate due to the
back-action noise of the detector, their coherence is preserved 
on the short time scales of order $\Omega^{-1}$ in the relevant 
case of weak interaction with the detector, $\Gamma \ll \Omega$. 
The short-time quantum coherence between energy eigenstates 
is constantly being created by the qubit-detector coupling that 
does not commute with the qubit Hamiltonian. The fact that it is 
incorrectly {\em neglected by assumption} in \cite{b1} leads to 
wrong results for the spectral density $S(\omega)$ at 
$\omega\simeq \Omega$. 

Quantitatively, the assumption of suppressed quantum coherence 
translates  into the form of the current (super)operator 
(Eq.~9 of \cite{b1}) written as a sum of several non-interfering
``jump processes'' in the energy domain. Since this description 
is adequate only on the long time scales, it produces correct 
expression for the current spectrum $S(\omega)$ at low frequencies 
$\omega \simeq \Gamma$ (see unnumbered equation after Eq. 13 in 
\cite{b1}) that coincides with the previous results \cite{b2,b5}, 
but does not reproduce current oscillations with frequency 
$\Omega$.

Correct expression for the current operator reduced to the 
qubit space can be obtained for the QPC detector if the bias 
voltage $V$ across the QPC is large, $eV\gg \Omega$, so that the 
detector and qubit dynamics have different characteristic times 
(see \cite{b2,b11}). The corresponding expression (e.g., Eq.~6 
in \cite{b11}) differs from Eq.~9 of \cite{b1} in that the 
different tunneling processes in the detector are added 
coherently, implying that proper ``unravelling'' of the qubit 
dynamics in conditional approach should be done as in 
\cite{b12,b3}, according to values of the detector current, not 
energy. This effectively assumes that there is no coherence 
between states with different number of electrons passed through 
the detector. For $eV \sim \Omega$, it is impossible to write 
down the current operator in the qubit space alone, since the 
qubit and detector dynamics cannot be separated. Perturbative 
treatment of the coupled qubit-detector dynamics \cite{b5} 
shows that, as expected, the oscillation peak in spectral 
density exists for sufficiently large bias voltages $eV>\Omega$.

\vspace{0.5cm}

\noindent
Dmitri V. Averin \\
\indent Department of Physics and Astronomy, USB, SUNY\\ 
\indent Stony Brook, NY 11794-3800

\noindent
Alexander N. Korotkov\\
\indent Department of Electrical Engineering, UC Riverside\\
\indent Riverside, CA 92521-0204


\begin{references}

\vspace{-1.5cm}

\bibitem{b1} T.M. Stace and S.D. Barrett, Phys.\ Rev.\ Lett. {\bf 92},
136802 (2004).

\bibitem{b2} A.N. Korotkov and D.V. Averin, Phys. Rev. B {\bf 64},
165310 (2001); A.N. Korotkov, Phys. Rev. B {\bf 63}, 085312 (2001);
D.V. Averin, in: {\em ``Exploring the quantum/classical frontier''}, 
Ed.\ by J.R. Friedman and S. Han, (Nova Science Publishes, NY, 2003), 
p.\ 447; cond-mat/0004364.

\bibitem{b3} H.S. Goan and G.J. Milburn, Phys.\ Rev. B {\bf 64},
235307 (2001).

\bibitem{b4} R. Ruskov and A.N. Korotkov, Phys. Rev. {\bf 67},
075303 (2003).

\bibitem{b5} A. Shnirman, D. Mozyrsky, and I. Martin, 
cond-mat/0311325.

\bibitem{b7} L.N. Bulaevskii and G. Ortiz, Phys. Rev. Lett.
{\bf 90}, 040401 (2003).

\bibitem{b8} Z. Nussinov, M.F. Crommie, and A.V.  Balatsky,
Phys.\ Rev. B {\bf 68}, 085402 (2003).

\bibitem{b9} E. Il'ichev {\it et al.}, Phys.\ Rev.\ Lett. {\bf 91}, 
097906 (2003). 

\bibitem{b11} W. Mao, D.V. Averin, R. Ruskov, and A.N. Korotkov,
cond-mat/0401484.

\bibitem{b12} A.N. Korotkov, Phys. Rev. B {\bf 60}, 5737 (1999).

\end{references}
\end{document}